\documentclass{article}
\usepackage{latexsym,amsmath,amssymb,amscd,amsfonts,amsthm,enumerate}

\begin{document}

\title{Singularity Analysis of a Variant of the Painlev{\'e}--Ince Equation}
\author{Amlan K Halder\thanks{%
Email: amlan.haldar@yahoo.com} \\
{\ \ \textit{Department of Mathematics, Pondicherry }}\\
{\ \textit{University, Kalapet, India-605014}}
\and Andronikos Paliathanasis\thanks{%
Email: anpaliat@phys.uoa.gr} \\
{\ \textit{}}
{\textit{}}
{\textit{Institute of Systems Science, Durban University of Technology}}\\
{\textit{Durban 4000, Republic of South Africa.}}\\
\and PGL Leach\thanks{%
Email: leachp@ukzn.ac.za} \\
{\textit{School of Mathematics, Statistics and Computer Science,}}\\
{\textit{University of KwaZulu-Natal, Durban, South Africa.}}\\
{\textit{Institute of Systems Science, Durban University of Technology}}\\
{\textit{Durban 4000, Republic of South Africa.}}}

\maketitle

\begin{abstract}
We examine by singularity analysis an equation derived by reduction using
Lie point symmetries from the Euler--Bernoulli Beam equation which is the
Painlev\'{e}--Ince Equation with additional terms. The equation possesses
the same leading-order behaviour and resonances as the Painlev\'{e}--Ince
Equation and has a Right Painlev\'{e} Series. However, it has no Left Painlev%
\'{e} Series. A conjecture for the existence of Left Painlev\'{e} Series for
ordinary differential equations is given.

\end{abstract}


\vspace{1.5cc}

\section{Introduction}

The Painlev\'{e}--Ince Equation
\begin{equation}
y^{\prime \prime }+3yy^{\prime}+y^{3}=0,  \label{1.1}
\end{equation}%
where the prime denotes differentiation of the dependent variable, $y(x)$,
with respect to the independent variable, $x$, is an equation noted for its
interesting properties. Firstly it possesses eight Lie point symmetries \cite%
{Mahomed 85 a} with the Lie Algebra $sl(2,R)$ which means that it is
linearisable by a point transformation. Secondly it is the second member of
the Riccati Hierarchy which is based upon the Riccati Equation with the
recursion operator $D+y$, where $D$ is the operator $\frac{d}{dy}$ \cite%
{Euler 07 a}. Thirdly it possesses the Painlev\'{e} Property in a very
unique way in that there are two possible Laurent expansions about its
simple pole. The coefficient of the leading-order term can be $1$ or $2$.
For the former the resonances are the generic $-1$ and $1$. For the latter
the resonances are again $-1$ and an unexpected $-2$. This meant that the
Laurent expansion must be decreasing in exponent from the simple pole \cite%
{Lemmer 93 a}. The existence of a resonance at an exponent lower than the
exponent of the leading-order term was not expected and the proposal was
derided by some, but the careful analysis of Feix and his team \cite{Feix 97
a, Feix 05 a, Claude 02 a} established the sense with the addition of the
Left Painlev\'{e} Series to the well-established Right Painlev\'{e} Series.
One notes that Andriopoulos \textit{et al } \cite{Andriopoulos 06 a}
discussed the existence of both positive and negative {nongeneric}
resonances for higher-order equations, \textit{ie} Mixed Painlev\'{e} Series
and their geometric interpretation.

In an analysis of the Euler--Bernoulli Equation for a beam Halder \textit{et
al} \cite{Halder 19 a} determined the Lie point symmetries \footnote{%
The Mathemaica add-on Sym \cite{Dimas 04 a, Dimas 06} was used to obtain the
symmetries.} and used them initially to reduce the fourth-order partial
differential equation to a fourth-order ordinary differential equation. This
fourth-order equation had sufficient Lie point symmetries to reduce it to a
second-order equation devoid of Lie point symmetries. It was noted that the
second-order equation was the Painlev\'{e}--Ince Equation with some rather
messy additional terms of order lower than the second order in the
derivative. It is the purpose of this paper to demonstrate that this
equation does in fact satisfy the requirement of the ARS algorithm \cite%
{Ablowitz 78 a, Ablowitz 80 a, Ablowitz 80 b}. The equation is nonautonomous
and so we treat it in the spirit of the treatment of nonautonomous equations
as found in Andriopoulos \textit{et al} \cite{Andriopoulos 11 a}.

\section{The equation and its singularity analysis}

The nonlinear second-order ordinary differential equation of our
consideration is
\begin{equation}
y^{\prime \prime }+y^{\prime }\left( \frac{7}{x}+3y\right) +y^{3}+\frac{%
7y^{2}}{x}+y\left( \frac{1}{16abx^{4}}+\frac{39}{4x^{2}}\right) +\frac{3}{%
2x^{3}}=0,  \label{2.1}
\end{equation}%
where the symbolic usages are the same as for (\ref{1.1}) above. As,
mentioned in the introduction, equation (\ref{2.1}) caught our attention, as
it is obtained from the reduction of the well-known Euler-Bernoulli Beam
equation, which is something new according to the author's knowledge and
cannot be found in the literature. A first glance at the equation establish
the fact that, it is a variant of the Painlev\'{e}--Ince equation. It has
zero Lie-point symmetries, and we choose to study the singularity analysis
to ascertain its integrability. A problem of interests also in the area of
physics where the Euler-Bernoulli Beam equation has applications. The
similarities and dissimilarities of equation (\ref{2.1}) with the Painlev%
\'{e}--Ince equation have been mentioned subsequently. \newline
To determine the leading-order behaviour of equation (\ref{2.1}), we
substitute
\begin{equation}
y=a(x-x_{0})^{p},  \label{2.2}
\end{equation}%
where $x_{0}$ is the location of the putative singularity and, being
movable, is one of the constants of integration to be determined from the
initial conditions, into (\ref{2.1}) to obtain
\begin{eqnarray}
0 &=&\frac{3x^{4}}{2}%
-apx^{7}(x-x_{0})^{-2+p}+ap^{2}x^{7}(x-x_{0})^{-2+p}+7apx^{6}(x-x_{0})^{-1+p}
\notag \\
&&+\frac{x^{3}(x-x_{0})^{p}}{16b}+\frac{39}{4}%
ax^{5}(x-x_{0})^{p}+7a^{2}x^{6}(x-x_{0})^{2p}+  \notag \\
&&+a^{3}x^{7}(x-x_{0})^{3p}+3a^{2}px^{7}(x-x_{0})^{-1+2p}.  \notag
\label{2.3}
\end{eqnarray}%
We consider the dominant terms to compute the value of $p$, which is,
\begin{equation*}
-apx^{7}(x-x_{0})^{-2+p}+ap^{2}x^{7}(x-x_{0})^{-2+p}+a^{3}x^{7}(x-x_{0})^{3p}+3a^{2}px^{7}(x-x_{0})^{-1+2p},
\end{equation*}%
we equate the exponents of the dominant terms, which are $-2+p,3p$ and $-1+2p
$, from which it is evident that $p=-1$, \textit{ie}, the singularity is a
simple pole.

We solve the dominant terms
\begin{equation}
\frac{2ax^{7}}{(x-x_{0})^{3}}-\frac{3a^{2}x^{7}}{(x-x_{0})^{3}}+\frac{%
a^{3}x^{7}}{(x-x_{0})^{3}}=0,  \label{2.4}
\end{equation}%
for $a$ and obtain $a=0,1,2$. So far all is the same as for (\ref{1.1}). To
determine the resonances for the leading-order coefficient $a=1$ we make the
substitution
\begin{equation}
y=(x-x_{0})^{-1}+m(x-x_{0})^{-1+s},  \label{2.5}
\end{equation}%
into (\ref{2.1}), take the coefficient of $m$ and require it to vanish for
the terms corresponding to the dominant terms from which $p$ was found.

The values of $s$ so found are $-1$ and $1$. The former value of $s$, which
is $-1$, is generic in nature and is related to the arbitrariness of the
location of the movable singularity. The details regarding the various
values of the resonances and most importantly the implications of the value $%
-1$ is explained in \cite{Andriopoulos 11 a}.

When we use the value $a = 2$, the resonances are $-1$ and $-2$. Thus far
the results are as for the Painlev{\'e}--Ince Equation, (\ref{1.1}).

To determine consistency we need to substitute a (truncated) Laurent
expansion into the full equation, which is,
\begin{equation*}
\frac{1}{(x-x_{0})}%
+a_{0}+a_{1}(x-x_{0})+a_{2}(x-x_{0})^{2}+a_{3}(x-x_{0})^{3},
\end{equation*}%
where, $a_{0},a_{1},a_{2}$ and $a_{3}$ are the coefficients of the series
terms. For the coefficient constants $a_{0},a_{1},a_{2}$ and $a_{3}$ in the
Laurent expansion we obtain the following results
\begin{eqnarray}
&&a_{1}=\frac{-1-156bx_{0}^{2}-224a_{0}bx_{0}^{3}-48a_{0}^{2}bx_{0}^{4}}{%
48bx_{0}^{4}},  \notag \\
&&a_{2}=\frac{%
11+2a_{0}x_{0}+1380bx_{0}^{2}+2104a_{0}bx_{0}^{3}+896a_{0}^{2}bx_{0}^{4}+128a_{0}^{3}bx_{0}^{5}%
}{128bx_{0}^{5}}\mbox{\rm and}  \notag \\
&&a_{3}=\frac{%
-(1+8bx_{0}^{2}(372+173a_{0}x_{0}+30a_{0}^{2}x_{0}^{2})+80b^{2}x_{0}^{4}(4095+...))%
}{11520b^{2}x_{0}^{8}}.  \notag
\end{eqnarray}

One could continue further, but the successive coefficients in the Right
Painlev{\'{e}} Series are expressible in terms of the two arbitrary
constants, $x_{0}$ and $a_{0}$, which is as they should be.

We turn now to the second possibility, $a=2$, for the value of the
coefficient of the leading-order term for which the resonances are $-1$ and $%
-2$. We substitute
\begin{equation}
y=2(x-x_{0})^{-1}+a_{2}(x-x_{0})^{-2}+a_{3}(x-x_{0})^{-3}+a_{4}(x-x_{0})^{-4}+a_{5}(x-x_{0})^{-5},
\label{2.9}
\end{equation}%
in equation (\ref{2.1}), to check the consistency by computing the
coefficients of the Laurent expansion.

An attempt to put various coefficients to zero fails immediately due to the
presence of the term, $x^{4}$, in (\ref{2.1}) for which there is no
compensating term in the posited Left Painlev{\'{e}} Series.

\section{Conclusion}

The Painlev\'{e}--Ince Equation is the first known instance of an ordinary
differential equation possessing both Left and Right Painlev\'{e} Series.
Equation (\ref{2.1}), obtained by reduction using Lie point symmetries from
the Euler--Bernoulli Beam equation, is like a Painlev\'{e}--Ince Equation
with some additional terms. These additional terms do not affect the first
two steps of the ARS algorithm as they are not part of the dominant terms in
(\ref{2.1}). However, when it comes to the test for consistency, these
additional terms must play a role.

The basis of the singularity analysis is the existence of a Laurent
expansion about a singularity. There are three possible forms for a
convergent Laurent expansion. One is an expansion from singularity in
increasing powers of $\left( x-x_{0}\right) $, where $x_{0}$ is the location
of the singularity in the complex plane. This is the well-known right-Painlev%
\'{e} series. Secondly the series can descend in powers from the singularity
and one has the left-Painlev`e series. Thirdly there can be an expansion
between two singularities which gives rise to the mixed-Painlev\'{e} series.
All of these are well illustrated in \cite{anlp}. One should bear in mind
that the existence of, say, a left-Painlev\'{e} series does suggest that the
series, of the nature of an asymptotic series, is convergent over the rest
of the complex plane. On the other hand the mixed series gives a starting
point for a series which cannot be convergent over the whole complex plane
and so it must terminate at the boundary determined by another singularity.

As far as the equation of our problem is concerned, for the Right Painlev%
\'{e} Series there is no serious effect apart from making the coefficients
of the terms in the series more complex. Nevertheless those coefficients are
expressible in terms of the two constants of integration required for the
second-order equation. We note that the location of the movable pole plays
an active role in the further coefficients as predicted in \cite%
{Andriopoulos 11 a}.

The collapse of the Left Painlev\'e Series comes as no surprise in
retrospect as the negative powers cannot possibly balance positive powers.
Indeed it suggests strongly that an equation with nondominant terms might
not be able to possess a Left Painlev\'e Series.\newline
\newline
\textbf{Conjecture:} For a differential equation to possess a Left
Painlev\'e Series all terms must be dominant.

\section*{Acknowledgements}

AKH expresses grateful thanks to UGC (India) NFSC, Award No.
F1-17.1/201718/RGNF-2017-18-SC-ORI-39488 for financial support and Late
Prof. K.M.Tamizhmani for the discussions which AKH had with him which formed
the basis of this work. PGLL acknowledges the support of the National
Research Foundation of South Africa, the University of KwaZulu-Natal and the
Durban University of Technology and thanks the Department of Mathematics,
Pondicherry University, for gracious hospitality.


\vspace{2cc}

\begin{thebibliography}{99}
\bibitem{Ablowitz 78 a} Ablowitz MJ, Ramani A \& Segur H (1978) Nonlinear
Evolution Equations and Ordinary Differential Equations of Painlev\'e Type
\textit{Lett Nuovo Cimento} \textbf{23} 333-337.

\bibitem{Ablowitz 80 a} Ablowitz MJ, Ramani A \& Segur H (1980) A connection
between nonlinear evolution equations and ordinary differential equations of
P-type I \textit{Journal of Mathematical Physics} \textbf{21} 715-721.

\bibitem{Ablowitz 80 b} Ablowitz MJ, Ramani A \& Segur H (1980) A connection
between nonlinear evolution equations and ordinary differential equations of
P-type II \textit{Journal of Mathematical Physics} \textbf{21} 1006-1015.

\bibitem{Andriopoulos 06 a} Andriopoulos K \& Leach PGL (2006) An
interpretation of the presence of both positive and negative nongeneric
resonances in the singularity analysis \textit{Physics Letters A} \textbf{359%
} 199-203.

\bibitem{Andriopoulos 11 a} Andriopoulos K \& Leach PGL (2011) Singularity
analysis for autonomous and nonautonomous differential equations \textit{%
Applicable Analysis and Discrete Mathematics} \textbf{5} 230-239.

\bibitem{Dimas 04 a} Dimas S \& Tsoubelis D (2004, October) SYM: A new
symmetry-finding package for Mathematica \textit{In Proceedings of the 10th
International Conference in Modern Group Analysis} (University of Cyprus
Press) 64-70.

\bibitem{Dimas 06} Dimas S \& Tsoubelis D (2006, June) A new
Mathematica-based program for solving overdetermined systems of PDEs \textit{%
In 8th International Mathematica Symposium}.

\bibitem{Euler 07 a} Euler M, Euler N \& Leach PGL (2007) The Riccati and
Ermakov-Pinney hierarchies \textit{Journal of Nonlinear Mathematical Physics}
\textbf{14} 290-310.

\bibitem{Feix 97 a} Feix MR, G\'eronimi C, Cair\'{o} L, Leach PGL, Lemmer RL
\& Bouquet S\'E (1997) On the singularity analysis of ordinary differential
equations invariant under time translation and rescaling \textit{Journal of
Physics A: Mathematical and General} \textbf{30} 7437-7461.

\bibitem{Feix 05 a} Feix MR, G\'eronimi C \& Leach PGL (2005) Properties of
some autonomous equations invariant under homogeneity symmetries \textit{%
Problems of Nonlinear Analysis in Engineering Systems} \textbf{11} 26-34.

\bibitem{Claude 02 a} G\'eronimi C, Leach PGL \& Feix MR (2002) Singularity
analysis and a function unifying the Painlev\'e and $\Psi$ series \textit{%
Journal of Nonlinear Mathematical Physics} \textbf{9} \textbf{Second
Supplement} 36-48.

\bibitem{Halder 19 a} Halder AK , Paliathansis K \& Leach PGL (2019) A
Complete Study Of Different Forms Of Beam Equations Through Lie Group
Analysis{submitted}.

\bibitem{Lemmer 93 a} Lemmer RL \& Leach PGL (1993) The Painlev\'{e} test,
hidden symmetries and the equation $y^{\prime\prime}+ yy^{\prime}+ky^{3} = 0$
\textit{Journal of Physics A: Mathematical and General} \textbf{26}
5017-5024.

\bibitem{Mahomed 85 a} Mahomed FM \& Leach PGL (1985) The linear symmetries
of a nonlinear differential equation \textit{Quaestiones Mathematicae}
\textbf{8} 241-274.

\bibitem{anlp} Paliathanasis A \& Leach PGL (2016) Nonlinear Ordinary
Differential Equations: A discussion on Symmetries and Singularities \textit{%
International Journal Geometric Methods Modern Physics} \textbf{13} 1630009
\end{thebibliography}
\end{document}